\documentclass[aps,twocolumn,showpacs,prl,amsmath,amssymb,floatfix,superscriptaddress]{revtex4-1}

\usepackage{color}
\usepackage{graphicx}
\usepackage{dcolumn}
\usepackage{bm}
\usepackage{array}
\usepackage{float}
\usepackage{supertabular}
\usepackage{longtable}
\usepackage{mathrsfs}
\usepackage{txfonts}
\usepackage[T1]{fontenc}
\usepackage[latin1]{inputenc}
\usepackage{amssymb}
\usepackage[usenames,dvipsnames]{xcolor}
\usepackage{amsmath}
\usepackage{amstext}
\usepackage{latexsym}
\usepackage[colorlinks=true,citecolor=Cerulean,linkcolor=RubineRed,urlcolor=Cerulean]{hyperref}

\def\vl{{{\lambda}}}

\def\la{\langle}
\def\ra{\rangle}
\def\om{\omega}

\def\tr{{\rm tr}}

\newcommand{\beq}{\begin{equation}}
\newcommand{\eeq}{\end{equation}}
\newcommand{\beqa}{\begin{eqnarray}}
\newcommand{\eeqa}{\end{eqnarray}}

\begin{document}
\title{Universal Work Fluctuations during Shortcuts To Adiabaticity by Counterdiabatic Driving}
\author{Ken Funo}
\affiliation{School of Physics, Peking University, Beijing 100871, China}
\author{Jing-Ning Zhang}
\affiliation{Center for Quantum Information, Institute for the Interdisciplinary Information Sciences, 
Tsinghua University, Beijing, 100084, P. R. China}
\author{Cyril Chatou}
\affiliation{Universit\'e Paris 13, Sorbonne Paris Cit\'e, 99 Avenue J.-B. Cl\'ement, F-93430 Villetaneuse, France}
\affiliation{Department of Physics, University of Massachusetts, Boston, MA 02125, USA}
\author{Kihwan Kim}
\affiliation{Center for Quantum Information, Institute for the Interdisciplinary Information Sciences, 
Tsinghua University, Beijing, 100084, P. R. China}
\author{Masahito Ueda}
\affiliation{Department of Physics, University of Tokyo, Hongo 7-3-1, Bunkyo-ku, Tokyo 113-0033, Japan}
\affiliation{RIKEN Center for Emergent Matter Science (CEMS), Wako, Saitama 351-0198, Japan}
\author{Adolfo del Campo}
\affiliation{Department of Physics, University of Massachusetts, Boston, MA 02125, USA}

\begin{abstract}
Counterdiabatic driving (CD) exploits auxiliary control fields to tailor the nonequilibrium dynamics of a quantum system, making possible the suppression of dissipated work in finite-time thermodynamics and the engineering of optimal thermal machines with no friction. We show that while the mean work done by  the auxiliary controls vanishes, CD leads to a broadening of the work distribution. We derive a fundamental inequality that relates nonequilibrium work fluctuations to the operation time and quantifies the thermodynamic cost of CD in both critical and noncritical systems.

\end{abstract}


\maketitle

Understanding the far-from-equilibrium dynamics of quantum systems is an open problem at the frontiers of physics. Yet, tailoring such dynamics is a necessity for the advancement of quantum technologies.
This challenge is fully embodied within the field of quantum thermodynamics with potential applications in energy science. Our understanding of nonequilibrium behavior of thermodynamic systems has deepened profoundly owing to fluctuation theorems (e.g., the Jarzynski equality) and stochastic thermodynamics~\cite{Jarzynski97,Crooks,Seifert,Tasaki,Kurchan,fluctuation1}. For example, the Jarzynski equality~\cite{Jarzynski97} has been used to find the equilibrium free energy of a system through measurements of the fluctuating nonequilibrium work~\cite{Jarexperiment}, and quite recently, its quantum version~\cite{Tasaki,Kurchan} has been tested experimentally in a trapped ion system~\cite{quantumjarexp}. In an effort to develop control tools to engineer the dynamics of thermal machines, schemes to  suppress excitations and work fluctuations  have been put forward~\cite{Minimalwork1,Minimalwork2,Wfluc1}. In any physical implementation, thermal machines such as  quantum heat engines and refrigerators must operate in a finite time to achieve a nonvanishing output power. This motivates the study of finite-time thermodynamics that targets the optimization of the trade-off between efficiency and power~\cite{SS07}. In this context, control techniques known as shortcuts to adiabaticity (STA) have emerged as a disruptive paradigm as they reproduce the quantum adiabatic dynamics of the system by suppressing excitations without the requirement of slow driving~\cite{STAR}.  STA have been used to boost the performance of quantum heat engines by enhancing its output power at zero friction~\cite{STA1,STA2,STA3} and to suppress work fluctuations, assisting, for example, the convergence of the Jarzynski equality~\cite{Wfluc1}. 

Assessing the cost of implementing STA arises as a natural question with both fundamental and practical implications in nonequilibrium statistical mechanics. Among the different techniques to engineer STA \cite{STAR}, counterdiabatic driving stands out as a unifying framework \cite{DR03,DR05,Berry09}. It  relies on the use of auxiliary control fields so that the exact evolution along STA is described by the adiabatic approximation to the dynamics of the (uncontrolled) system, even in arbitrarily fast processes. Since the introduction of STA, it is understood that the amplitude of the auxiliary control field increases as the duration of the STA is reduced \cite{Demirplak08,Santos,DRZ12,Zheng15,CD16}. 
In this Letter, we elucidate the thermodynamic cost of counterdiabatic driving by studying how work fluctuations are modified during STA.  We show that  the work done by the counterdiabatic fields vanishes on average. However, we find that STA modify the work probability distribution and increase work fluctuations, whose amplitude is set by the quantum geometric tensor of the underlying Hilbert space. Furthermore, we derive a fundamental nonequilibrium inequality that relates work fluctuations to the duration of the process.

\noindent {\it Shortcuts to Adiabaticity by Counterdiabatic Driving.---}\ Consider a time-dependent Hamiltonian $\hat{H}_0 (\vl_{t})$ with instantaneous eigenvalues $\{\varepsilon_n(t)\}$ and eigenstates $\{|n(\vl_{t})\ra\}$, depending explicitly on a set of parameters $\vl_{t}=(\lambda^1 (t),\dots,\lambda^N (t))$. Here, we fix the initial and final parameters to $\lambda_{0}=\lambda_{\mathrm{i}}$ and $\lambda_{\tau}=\lambda_{\mathrm{f}}$, respectively, where $\tau$ is the time required to complete the protocol. We pose the problem of driving an initial state $|n(\lambda_{\mathrm{i}})\ra$ to a final state in a given finite time $\tau$, so that the final state matches $|n(\lambda_{\mathrm{f}})\ra$. In the following, we simplify the notation of this protocol dependence, e.g., $|n(\vl_{t})\ra=|n(t)\ra$.  
A technique which achieves this goal is the so-called counterdiabatic driving (CD),  also known as transitionless quantum driving~\cite{DR03,Berry09}, that will be our focus in this Letter.
CD has inspired several experiments \cite{expCD1,expCD2} and has recently been implemented in both discrete and continuous-variable systems \cite{expCD3,expCD4}. 
Given a protocol $\vl_{t}$, whenever $\hat{H}_0 (t)$ is slowly-varying, the dynamics of the $n$-th eigenstate $|n(0)\ra$ in the adiabatic approximation at time $t$ reads
\beq
\label{adiabsol}
|\chi_n(t)\ra=e^{-\frac{i}{\hbar}\int_0^tdt'\varepsilon_n(t')-\int_0^tdt'\frac{d{\lambda}^\mu}{dt'}\la n(t')|\partial_{\mu}n(t')\ra }|n(t)\ra,
\eeq
where $\partial_{\mu}=\frac{\partial}{\partial \lambda^\mu}$, and summation over  repeated Greek indices is implicit. The first and second terms in the exponent of Eq.~(\ref{adiabsol}) correspond to the dynamical and geometric phase, respectively. The primary goal  of CD is to find a Hamiltonian $\hat{H}_{\rm CD}$ for which the adiabatic approximation to $\hat{H}_{0}(t)$ 
becomes the exact solution of the time-dependent Schr\"odinger equation for $\hat{H}_{\rm CD}(t)$, i.e., $\hat{U}_{\rm CD}(t,0)|n(0)\ra=|\chi_{n}(t)\ra$, where $\hat{U}_{\rm CD}(t,0):=\mathrm{T}\exp[-\frac{i}{\hbar}\int^{t}_{0}dt' \hat{H}_{\rm CD}(t')]$ is the time evolution operator and $\mathrm{T}$ denotes the time-ordering operator. Direct construction of the time-evolution operator 
\beqa
\hat{U}_{\rm CD}(t,0)=\sum_n |\chi_n(t)\ra\la n(0)|,
\eeqa
yields an explicit form of $\hat{H}_{\rm CD}(t)$  \cite{DR03,DR05,Berry09}, 
\beqa
\label{hCD}
& &\hat{H}_{\rm CD}(t)=i\hbar\left[\partial_t\hat{U}_{\rm CD}(t,0)\right]\hat{U}^{\dag}_{\rm CD}(t,0)=\hat{H}_0 (t)+\hat{H}_1 (t),\\
& &\hat{H}_1 (t)= i\hbar\sum_n(|\partial_tn(t)\ra\la n(t)|-\la n(t)|\partial_t n(t)\ra |n(t)\ra\la n(t)|).\nonumber 
\eeqa
We find that $\hat{H}_{1}(t)$ 
is the auxiliary term required to drive the system from $|n(0)\ra$ to $|n(\tau)\ra$  for all $n$ in a given time $\tau$, maintaining adiabaticity with respect to $\hat{H}_0$. As a result, the CD control Hamiltonian $\hat{H}_{1}(t)$ differs from similar Hamiltonians that appear in the proof of the adiabatic theorem \cite{Avron87}.
We shall assume that the auxiliary term is switched off at the initial and final stages of the process, i.e., $\hat{H}_{1}(0)=\hat{H}_{1}(\tau)=0$. 
We note that the evolution $|\chi_n(t)\ra$ is non-adiabatic with respect to the full driving Hamiltonian, whose instantaneous eigenstates satisfy
\beqa
\hat{H}_{\rm CD}(t)|\Psi_n(t)\rangle=E_n(t)|\Psi_n(t)\rangle.
\eeqa

\noindent {\it Work Fluctuations under Counterdiabatic Driving.---} We next study how work fluctuations along CD are modified with respect to a truly adiabatic process (i.e., the limit of slow driving when $\hat{H}_{1}(t)\rightarrow 0$). To do so, we introduce a work cost of the system for a microscopic trajectory of the system. Suppose that we start from the canonical distribution whose occupation probability in the energy eigenstate $|n(0)\ra$ is 
$p_{n}^{0}=\exp[-\beta\varepsilon_{n}(0)]/Z$, where $Z=\sum_n\exp[-\beta\varepsilon_n(0)]$ is the partition function. If we observe a trajectory starting from $|n(0)\ra$ and find the state of the system being $| \Psi_k(t)\ra$ at time $t$, the probability of obtaining that trajectory is given by
\beq
p_n^0 p_{n\rightarrow k}^t
:=p_n^0 |\la \Psi_k(t)|\hat{U}_{\rm CD}(t,0)|n(0)\ra|^2
=p_n^0 |\la \Psi_k(t)|n(t)\ra|^2,\label{jointprob}
\eeq
and the work cost along that trajectory is given by $E_{k}(t)-\epsilon_{n}(0)$. Note that we need to perform two energy measurements at times $t'=0$ and $t$ to obtain the probability distribution~(\ref{jointprob}), which is referred to as the two-point measurement scheme~\cite{Tasaki,Kurchan,fluctuation1}. The explicit expression for the work probability distribution $P[W(t)]$ reads
\beqa
\label{P(W)}
P[W(t)]:=\sum_{k,n}p_n^0p_{n\rightarrow k}^t\delta[W(t)-(E_k(t)-\varepsilon_n(0))].
\eeqa

In the truly adiabatic limit, $H_1$ vanishes and $\hat{H}_{\rm CD}(t)=\hat{H}_0(t)$ for all $t$, and  the transition probability becomes the Kronecker delta: $p_{n\rightarrow k}^t=\delta_{k,n}$. 
As a result, the work probability distribution takes the form
\beqa
P_{\rm ad}[W(t)]
=\sum_{n}p_n^0\delta[W(t)-W_{\rm ad}^{(n)}(t)],
\eeqa
where $W_{\rm ad}^{(n)}(t):=\epsilon_{n}(t)-\epsilon_{n}(0)$ is the work cost along the adiabatic trajectory. 

Because $\hat{H}_{\mathrm{CD}}(\tau)=\hat{H}_{0}(\tau)$
, we obtain $p^{\tau}_{n\rightarrow k}=|\la k(\tau)|n(\tau)\ra|^{2}=\delta_{k,n}$ and thus $P[W(\tau)]=P_{\rm ad}[W(\tau)]$. 
In particular, an initial thermal state $\rho(0)=\sum_np_n^0|n(0)\ra\la n(0)|$ evolves into $\rho(\tau)=\sum_np_n^0|n(\tau)\ra\la n(\tau)|$. 
Therefore, at the end of the protocol, all properties about $W$ for CD become equivalent to those for the adiabatic dynamics. 
However, $P[W(t)]$ and $P_{\rm ad}[W(t)]$ are different at the intermediate stage. In what follows, we analyze the deviations of the mean and variance of work along CD from those of the adiabatic dynamics for arbitrary $0\leq t\leq \tau$. 

We first show that the average work cost along the CD evolution is always equal to the adiabatic value,
\beqa
\label{equality1}
\la W(t)\ra=\la W(t)\ra_{\rm ad}.
\eeqa
We note that from the instantaneous Schr\"odinger equation for $H_{\rm CD}(t)$ it is possible to derive
\beqa
\label{keyeq}
(E_m (t)-\varepsilon_n (t))\la n(t)|\Psi_m(t)\ra
&=&\la n(t)|H_1|\Psi_m(t)\ra. 
\eeqa
An explicit evaluation of the right-hand side leads to
\beqa
\la n(t)|H_1|\Psi_m(t)\ra&=&i\hbar \dot{\lambda}^\mu\sum_{k(\neq n)}\la n(t)|\partial_{\mu}k(t)\ra\la k(t)|\Psi_m(t)\ra.\nonumber
\eeqa
 Multiplying both sides of this equation by $\la \Psi_m(t)|n(t)\ra$ and summing over the quantum number $m$ labeling the instantaneous eigenstate of the full driving Hamiltonian, one finds
\beq
\sum_mp_{n\rightarrow m}^t(E_m (t)-\varepsilon_n (t))=0,\label{keyeq3}
\eeq
and in particular,
\beq
\sum_{n,m}p_n^0p_{n\rightarrow m}^t[(E_m (t)-E_n (0))-(\varepsilon_n (t)-\varepsilon_n (0))]=0, \label{keyeq1}
\eeq
which proves the equality between the mean work under CD and the adiabatic work, i.e.,  Eq.~(\ref{equality1}).
 Therefore, we find no difference between CD and the adiabatic dynamics in terms of the mean work. However, as we shall see below, CD alters work fluctuations with respect to the adiabatic dynamics, and we identify this difference as the thermodynamic cost to implement CD.

We next characterize the CD work fluctuations with respect to the adiabatic trajectory. 
Taking the absolute square of (\ref{keyeq}), multiplying $p^{0}_{n}$ and  summing over $n$ and $m$, we have
\beqa
\sum_{n,m}p^{0}_{n}p_{n\rightarrow m}^t (E_m (t)-\varepsilon_n (t))^2&=&\hbar^2  \sum_{n}p^{0}_{n}\la \partial_{\mu}n(t)|P_n^\perp|\partial_{\nu}n(t)\ra \dot{\lambda}^\mu\dot{\lambda}^\nu \nonumber \\
&=&\hbar^2  \sum_np_n^0 g_{\mu\nu}^{(n)}\dot{\lambda}^\mu\dot{\lambda}^\nu , \label{keyeq2}
\eeqa
where $P_n^\perp=1-P_n$ is the projector onto the space orthogonal to that spanned by the state $|n(t)\ra$ with  $P_n=|n(t)\ra\la n(t)|$ being the projector on $|n(t)\ra$. In Eq.~(\ref{keyeq2}), we have identified the metric $g_{\mu\nu}^{(n)}={\rm Re}Q_{\mu\nu}^{(n)}$ with the real part of the quantum geometric tensor of the $|n(t)\ra$-state manifold introduced by Provost and Vallee \cite{PV80},
\beqa
Q_{\mu\nu}^{(n)}:=\la \partial_{\mu}n(t)|P_n^\perp|\partial_{\nu}n(t)\ra. \label{metric}
\eeqa
Note that Eq.~(\ref{metric}) dictates the quadratic decay of the square-root of the fidelity between two states $|n(t)\ra$ and 
$|n(t+\delta t)\ra$, i.e., $\sqrt{F(|n(t)\ra,|n(t+\delta t)\ra)}:=|\la n(t)|n(t+\delta t)\ra|=1-g_{\mu\nu}^{(n)}\dot{\lambda}^{\mu}\dot\lambda^{\nu}\delta t^{2}/2+\mathcal{O}(\delta t^3)$. Multiplying by $2(\varepsilon_{n}(t)-\varepsilon_{n}(0))$  Eq.~(\ref{keyeq3}), we have 
\beq
2\sum_{n,m}p^{0}_{n}p_{n\rightarrow m}^t (\varepsilon_{n}(t)-\varepsilon_{n}(0))(E_{m}(t)-\varepsilon_{n}(t))=0.
\eeq
Adding the above equation to the left-hand side of Eq.~(\ref{keyeq2}), we obtain
\beqa
\text{LHS}&=&\sum_{n,m}p^{0}_{n}p_{n\rightarrow m}^t \biggl[(E_m (t)-\varepsilon_n (0))^2-(\varepsilon_n (t)-\varepsilon_n (0))^2\biggr]\nonumber \\
&=&\la W^{2}(t)\ra-\la W^{2}(t)\ra_{\rm ad}.\label{keyeq4}
\eeqa
By combining Eqs.~(\ref{equality1}),  (\ref{keyeq2})  and (\ref{keyeq4}), we find that 
\beq
\delta (\Delta W)^{2}:=\mathrm{Var}[W(t)]-\mathrm{Var}[W(t)]_{\rm ad}
=\hbar^2  \sum_np_n^0g_{\mu\nu}^{(n)}\dot{\lambda}^\mu\dot{\lambda}^\nu ,
\label{wvarq}
\eeq
where $\text{Var}[W]:=\int\mathrm{d}W\hspace{.1mm}P(W) W^{2}-(\int\mathrm{d}W\hspace{.1mm}P(W)W)^{2}$. This result is remarkable in that it relates the instantaneous  excess of  work fluctuations $\delta (\Delta W)^{2}$ with an exclusively geometric quantity, namely, the quantum geometric tensor $g_{\mu\nu}^{(n)}$ induced by the protocol $\lambda(t)$. The excess of work fluctuations quantifies how the work probability distribution along the protocol at time $t$ is broadened with respect to the adiabatic dynamics as a result of the CD protocol.

We define 
\beqa
\label{defL}
\ell(\rho(0),\rho(\tau))
:=\int^{\lambda_{\tau}}_{\lambda_{0}}\sqrt{\sum_{n}p_n^0 g_{\mu\nu}^{(n)}d{\lambda}^\mu d{\lambda}^{\nu}},
\eeqa
as a natural distance between $\rho(0)$ and $\rho(\tau)$ under the CD dynamics, that enforces parallel transport for each eigenmode independently.  Combining (\ref{wvarq}) and (\ref{defL}), we obtain
\beq
\label{qsl1}
\tau= \frac{\hbar \ell(\rho(0),\rho(\tau))}{\la \delta\Delta W \ra_{\tau}},
\eeq
where $\left\la f\right\ra_{\tau}:=\tau^{-1}\int^{\tau}_{0}f(t)dt$ denotes the time-average and $\delta \Delta W:=\sqrt{\delta (\Delta W)^{2}}$. 
As we show in~(\ref{supp:ineq}), we can further lower-bound $\ell(\rho(0),\rho(\tau))$ as
\beqa
\ell(\rho(0),\rho(\tau))
\geq  \mathcal{L}(\rho(0),\rho(\tau)),\label{lengthineq}
\label{inqls}
 \eeqa
 in terms  of the Bures length between two mixed states $\mathcal{L}(\rho,\sigma):= \arccos\sqrt{F(\rho,\sigma)}$, where $F(\rho,\sigma):=(\tr[(\hspace{-0.3mm}\sqrt{\rho}\hspace{0.8mm} \sigma\hspace{-0.5mm}\sqrt{\rho})^{1/2}])^2$ is the fidelity~\cite{Uhlmann92,Jozsa94}. On the other hand, as we show in~(\ref{supp:we}), we can upper bound the excess of the work fluctuations~(\ref{wvarq}) as follows:
\beq
 \delta (\Delta W)^{2} \leq (\Delta E_{\rm CD})^{2}:=\la H_{\rm CD}^{2}\ra-\la H_{\rm CD}\ra^{2}. \label{hvar}
\eeq
Combining Eqs.~(\ref{qsl1}) (\ref{inqls}) and (\ref{hvar}), we obtain
\beq
\tau\geq \frac{\hbar\mathcal{L}(\rho(0),\rho(\tau))}{\left\la \delta \Delta W\right\ra_{\tau}}\geq  \frac{\hbar \mathcal{L}(\rho(0),\rho(\tau))}{\la \Delta E_{\rm CD} \ra_{\tau}} .\label{qsl2}
\eeq
This result gives a tighter bound on the duration of time $\tau$ compared to the Mandelstam-Tamm time-energy uncertainty relation \cite{MT45, AA90,Uhlmann92,BC94}, i.e., $\tau\geq \hbar\mathcal{L}(\rho(0),\rho(\tau))/\la \Delta E\ra_{\tau}$, with the role of the time-averaged standard deviation of energy $\la \Delta E\ra_{\tau}$ replaced by the time-averaged excess of the work fluctuations  $\la \delta \Delta W\ra_{\tau}$. This quantity $\la \delta \Delta W\ra_{\tau}$ captures the enhancement of the fluctuations in work done on or by the system as the duration of the protocol is shortened. We identify this quantity, dictated by the geometry of Hilbert space, as the thermodynamic cost to implement CD. Therefore, inequality~(\ref{qsl2}) gives a novel type of the quantum speed limit that provides a quantification of the thermodynamic cost of shortening the protocol time $\tau$ of STA assisted by CD, that entails an  increase in the excess of work fluctuations with respect to the adiabatic limit.

We note that in the zero temperature limit and when $\lambda_{t}$ follows the geodesic connecting the initial and final states, the distance $\ell$ reduces to the Bures length for pure states~(\ref{supp:zerot}), i.e., $\ell(|\psi(0)\ra,|\psi(\tau\ra)=\mathcal{L}(|\psi(0)\ra,|\psi(\tau\ra)$ with $|\psi\ra$ denoting the ground state. As a result, equality is achieved in~(\ref{qsl2}) for pure states, i.e.,
\beq
\tau=\frac{\hbar \mathcal{L}(|\psi(0)\ra,|\psi(\tau\ra)}{\la \delta\Delta W\ra_\tau}.
\eeq

\begin{figure}[tbp]
\begin{center}
\includegraphics[width=.95\linewidth]{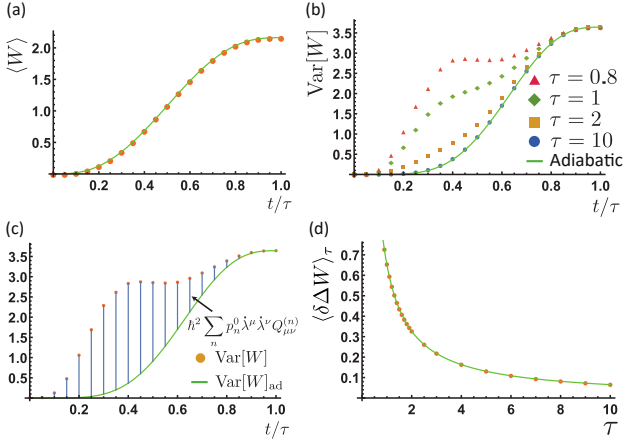}
\caption{{\bf Thermodynamic cost of counterdiabatic driving.} Numerical calculation for a time-dependent harmonic oscillator. (a) Instantaneous value of the average work cost. Orange dots are obtained via CD for $\tau=0.8$ and the green curve is obtained from the adiabatic protocol. (b) Time dependence of the variance of work for different values of the duration $\tau$ of the process. (c) Time dependence of each term in Eq.~(\ref{wvarq}) for $\tau=0.8$. (d) Time average of the work fluctuations $\left\la \delta \Delta W\right\ra_{\tau}$. The green curve is a $\tau^{-1}$ fit with its coefficient found to be $0.653$, which is larger than the Bures length $\mathcal{L}=0.476$ in accordance with~(\ref{qsl2}). }
\label{fig:TDHO3}
\end{center}
\end{figure}

\noindent {\it Trapped-ion implementation.---} We next consider a driven harmonic oscillator that can be implemented by a laser-induced potential in a trapped-ion system. Specifically,  we consider is a single $^{171}{\rm Yb}^+$ ion confined in a linear Paul trap, which is a natural platform for the experimental investigation of nonequilibrium work fluctuations and STA~\cite{quantumjarexp,expCD4}. In the interaction picture with respect to the quantum harmonic oscillator $\hat H_{\rm HO}=\hbar\left(\omega_0-\nu\right)\left(\hat a^\dag\hat a+\frac{1}{2}\right)$, the effective Hamiltonian of two distinct Raman processes (see Appendix {\it Ion-trap realization}) can be written as
\begin{eqnarray}
\hat H_{\rm eff}\left(t\right)&=&\hbar\nu\left(\hat a^\dag\hat a+\frac{1}{2}\right)+\frac{\hbar}{2}\left(\Omega_{\rm eff, 1}\left(t\right)\hat a \hat a+\Omega^{*}_{\rm eff, 1}\left(t\right)\hat a^\dag\hat a^\dag\right)\nonumber \\
&-&\frac{\hbar\Omega_{\rm eff, 2}\left(t\right)}{2}\left(\hat a\hat a^\dag+\hat a^\dag \hat a\right),
\end{eqnarray}
where $\nu$ is the detuning to the two-photon sideband Raman transitions and the annihilation operator $\hat a=\sqrt{m\omega_0/(2\hbar)}(\hat q+i\hat p/(m\omega_0))$ is defined on the motional mode with the (time-independent) trap frequency $\omega_0$, and a similar relation holds for the creation operator $\hat{a}^{\dagger}$.
The strengths of the laser-induced potentials are characterized by $\Omega_{\rm eff, 1}\left(t\right)$ and $\Omega_{\rm eff, 2}\left(t\right)$, which can be controlled by tuning the strengths and phases of the Raman laser beams. Note that $\Omega_{\rm eff, 2}\left(t\right)$ is real and $\Omega_{\rm eff,1}\left(t\right)$ can be complex. Setting $\Omega_{\rm eff, 1}\left(t\right)=-\Omega\left(t\right)+\frac{i\dot\omega\left(t\right)}{2\omega\left(t\right)}$ and $\Omega_{\rm eff, 2}\left(t\right)=\Omega\left(t\right)$ with the time-dependent frequency $\omega\left(t\right):=\sqrt{\nu\left(\nu-2\Omega\left(t\right)\right)}$, the above effective Hamiltonian can be rewritten as a sum of the quantum harmonic-oscillator term $\hat{H}_{0}(t)$ with the effective mass $m_{\rm eff}=m\omega_0/\nu$ and the auxiliary counterdiabatic field $\hat{H}_{1}(t)$~\cite{Muga10,delcampo13,DJD14} as follows:
\begin{eqnarray}
\hat{H}_{\rm eff}(t)&=&\hat{H}_{0}(t)+\hat{H}_{1}(t), \\
\hat H_{0}\left(t\right)&=&\frac{\hat p^2}{2m_{\rm eff}}+\frac{1}{2}m_{\rm eff}\omega\left(t\right)^2\hat q^2, \ \  \hat{H}_1\left(t\right)=-\frac{\dot{\om}}{4\om}(\hat q\hat p+\hat p\hat q). \nonumber 
\end{eqnarray}
The condition $\hat{H}_{1}=0$ at $t=\{0,\tau\}$ leads to the boundary conditions on $\om(t)$
that can be satisfied by  a polynomial ansatz, such as
$\om(t)=\om_{\mathrm{i}} +10\delta s^3-15\delta s^4+6\delta s^5$, where  $\om_{\mathrm{f}}=\om_{\mathrm{i}} +\delta$ and $s=t/\tau$. We use this driving protocol to analyze the thermodynamic cost of STA engineered via CD.
We numerically calculate $\left\la W(t)\right\ra$, $\mathrm{Var}[W(t)]$ and $\left\la \delta \Delta W\right\ra_{\tau}$ as functions of the evolution time $0\leq t\leq \tau$ along STA. The results are shown in Fig.~\ref{fig:TDHO3}. In the numerical calculation, we choose $\omega_{\mathrm{i}}=1$, $\omega_{\mathrm{f}}=3$ and set $m=\hbar=\beta=1$.
The mean work along STA is shown to match exactly the adiabatic value, in Fig.~\ref{fig:TDHO3} (a) in agreement with Eq.~(\ref{equality1}). As a result, the effect of the fast driving along STA is observed only on the work fluctuations and leads to a broadening of the work probability distribution $P[W(t)]$. The instantaneous variance of $W(t)$ surpasses the adiabatic value by a  quantity which is directly related to the quantum geometric tensor as shown in Fig.~\ref{fig:TDHO3} (b-c). Furthermore, the time-average work fluctuations with respect to the adiabatic trajectory are  bounded by the Bures length between the initial and final states, $\mathcal{L}(\rho(0),\rho(\tau))$, as dictated by (\ref{qsl2}). It follows that the thermodynamic cost of implementing the CD scheme is constrained by the geometry of the Hilbert space. This imposes a fundamental work-time uncertainty relation which determines the scaling of work fluctuations with respect to the duration of the process, as shown in Fig.~\ref{fig:TDHO3} (d).

{\it Quantum critical systems.---}
Equations (\ref{wvarq}) and (\ref{qsl2}) have direct implications on the work fluctuations  of many-body quantum  systems that exhibit quantum phase transitions by varying a parameter $\lambda$ of the uncontrolled system Hamiltonian $\hat{H}_0 (\lambda)$ across a critical value $\lambda_c$ at which the energy gap between the ground and first-excited state closes. The CD driving can be applied to this situation \cite{DRZ12,Takahashi13,Saberi14,Damski14,Campbell15}.
In the neighborhood of the critical point, the emergent conformal symmetry leads to the  divergence of the equilibrium correlation length  $\xi=\xi_0/|\lambda-\lambda_c|^\nu$, where $\nu$ is the correlation-length  critical exponent. This power-law behavior sets the scaling of the quantum geometric tensor \cite{Gu10}.  As a result, during the STA dynamics induced by CD, the work fluctuations exhibit a universal scaling that can be characterized by both the proximity to the critical point and the system size.
While the mean work done by the CD term remains equal to the adiabatic case, work fluctuations diverge in the thermodynamic limit. In particular, when the behavior is dominated by the low-energy excitations, we obtain
\beqa
\label{varWscaling}
\delta (\Delta W)^{2}\sim \frac{N}{|\lambda-\lambda_c|^{2-\nu d}},
\eeqa
where $N$ denotes the number of particles and $d$ is the dimension of the system.
At the critical point, the scaling with the system size reads
$\mathrm{Var}[W(\lambda_c)]-\mathrm{Var}[W(\lambda_c)]_{\rm ad}\sim N^{2/d\nu}$.
For the sake of illustration, we consider the quantum Ising chain, which is a prototypical model for quantum phase transitions described by the Hamiltonian 
\beqa
\hat{H}_0 [\lambda]= -\sum_{n=1}^N(\sigma_n^x\sigma_{n+1}^x+\lambda\sigma_n^z),
\eeqa
where $\lambda$ represents an external magnetic field and we consider periodic boundary conditions $\sigma_{N+1}^{x,z}=\sigma_{1}^{x,z}$.  The competition between the two terms in the Hamiltonian leads to a well-known quantum phase transition with  $\lambda_c=\pm1$ between a paramagnetic phase ($|\lambda|>1$) and a doubly degenerate ferromagnetic phase ($|\lambda|<1$). 
The relevant diagonal elements of the quantum geometric tensor for $\hat{H}_0 [\lambda]$ have recently been reported in closed form \cite{DR14}.
\begin{figure}[tbp]
\begin{center}
\includegraphics[width=.95\linewidth]{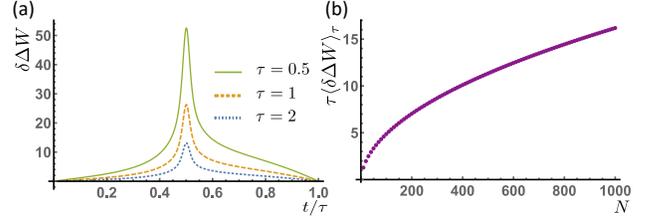}
\caption{{\bf Work fluctuations across a quantum phase transition.} 
(a) Instantaneous work fluctuations during the CD evolution of a quantum Ising chain as a function of time for different choices of the duration of the protocol $\tau=0.5,1,2$ with $\delta=1$. (b) Divergence of the time-integrated work fluctuations with the system size $N$.}
\label{fig2}
\end{center}
\end{figure}
We consider the symmetric counterdiabatic driving of the chain, initialized in its ground state, across the critical point $\lambda_c=1$ with an arbitrary protocol 
$\lambda(t)$ 
 satisfying $\dot{\lambda}=0$ at $t=0,\tau$, and amplitude $2\delta=|\lambda(\tau)-\lambda(0)|$.
For example, we take $\lambda(t)=1+\delta-6\delta (t/\tau)^2 +4\delta (t/\tau)^3$ but the following results hold independently of the concrete form of $\lambda(t)$.  During the evolution, the instantaneous work fluctuations shown in Fig.~\ref{fig2} exhibit a pronounced peak in the neighborhood of the critical point, $\lambda\approx \lambda_c$, in agreement with Eq. (\ref{varWscaling}). This scaling leads to a divergence of the time-integrated work fluctuation with the system size $N$, i.e., $\tau\left\la \delta \Delta W\right\ra_{\tau}\sim N^{\alpha}$. A fit to the numerical data leads with the power-law exponent $\alpha =0.516$ which is consistent with the theoretical value  $1/2$. The difference can be attributed to the fact that precisely at the critical point $\lambda=\lambda_c$  the scaling becomes linear in $N$ with $\nu=d=1$.

\noindent {\it Conclusion.---}
Shortcuts to adiabaticity have recently been proposed as a disruptive paradigm in finite-time thermodynamics to engineer thermal machines that operate at maximum-efficiency (zero friction) and arbitrary output power. We have analyzed the thermodynamic cost of  implementing the counterdiabatic driving scheme that provides a unifying framework to engineer such shortcuts.  In particular, we have shown that the mean work done by the auxiliary counterdiabatic fields vanishes~(\ref{equality1}), while the work fluctuations are substantially modified~(\ref{wvarq}). We have derived a fundamental inequality that constraints the enhancement of work fluctuations as a function of the duration of the process~(\ref{qsl2}) and proposed a trapped-ion test to verify our findings. 
Our work should find broad applications in quantum thermodynamics with applications in energy science and the advancement of trapped-ion quantum technology as a testbed for nonequilibrium statistical mechanics.

\begin{acknowledgments}
{\it Acknowledgements.---}
It is a pleasure to acknowledge illuminating discussions with Mathieu Beau. 
We further acknowledge funding support by UMass Boston (project P20150000029279), the John Templeton Foundation, the National Key Research and Development Program of China under Grants No. 2016YFA0301900 (No. 2016YFA0301901), and the National Natural Science Foundation of China (Grants No. 11374178, No. 11375012, No. 11534002, No. 11574002, and No. 11504197). This work was initiated during a visit of K.F. at the University of Massachusetts Boston supported by JSPS (Grant No. 254105), and we thank for its hospitality. M.U. acknowledges support from KAKENHI Grant No. 26287088 from the Japan Society for the Promotion of Science, a Grant-in-Aid for Scientific Research on Innovative Areas `Topological Materials Science' (KAKENHI Grant No. 15H05855), the Photon Frontier Network Program from MEXT of Japan, and the Mitsubishi Foundation. 

\end{acknowledgments}

\newpage

\appendix
 
 \section{Measures of distance between mixed states}
In this section, we prove that
\beqa
 \mathcal{L}(\rho(0),\rho(\tau))\leq \ell(\rho(0),\rho(\tau)). \label{supp:ineq}
\eeqa
To this end, we first introduce the fidelity for mixed states~\cite{Uhlmann92,Jozsa94},
\beq
F(\rho,\sigma):=(\tr[(\hspace{-0.3mm}\sqrt{\rho}\hspace{0.8mm} \sigma\hspace{-0.5mm}\sqrt{\rho})^{1/2}])^2.
\eeq
The fidelity for two neighboring mixed states reads
\beq
F(\rho(\lambda_{t}),\rho(\lambda_{t+dt}))=1-\frac{1}{2}\eta_{\mu\nu}\dot{\vl}^{\mu}\dot{\vl}^{\nu}dt^{2}+O(dt^{3}),
\eeq
where $\eta_{\mu\nu}$ is a Riemannian metric defined by~\cite{BC94}
\beq
\eta_{\mu\nu}=\frac{1}{4}\sum_{n}\frac{\partial_{\mu}p_{n}\partial_{\nu}p_{n}}{p_{n}}+\frac{1}{2}\sum_{k\neq n}\frac{(p_{n}-p_{k})^{2}}{p_{n}+p_{k}}\langle \partial_{\mu}n|k\rangle\langle k|\partial_{\nu}n\rangle  .
\eeq
Here $p_{n}$ and $|n\rangle$ are defined through the spectral decomposition of the density operator $\rho(t)=\sum_{k}p_{n}(t)|n(t)\rangle\langle n(t)|$. Note that under counterdiabatic driving, $p_{n}(t)=p^{0}_{n}$ is a constant, i.e., $\partial_{\mu}p^{0}_{n}=0$.

By introducing a distance using the metric $\eta_{\mu\nu}$ as
\beq
ds^{2}=\eta_{\mu\nu}d\vl^{\mu}d\vl^{\nu},
\eeq
the length of a curve obtained from a path followed by $\rho(t)$ satisfies
\beqa
\mathcal{L}(\rho,\sigma)\leq \int^{\tau}_{0}ds=\int_{\lambda_{0}}^{\lambda_{\tau}}\sqrt{\eta_{\mu\nu}d\vl^{\mu}d\vl^{\nu}},\label{supp:brues}
\eeqa
where 
\beqa
\mathcal{L}(\rho,\sigma):= \arccos\sqrt{F(\rho,\sigma)}
\eeqa
is the Bures distance~\cite{Uhlmann95}. Using the inequality
\beq
 \frac{(p^{0}_{n}-p^{0}_{k})^{2}}{p^{0}_{n}+p^{0}_{k}}\leq p^{0}_{n}+p^{0}_{k},
\eeq
we can bound from above the metric $\eta_{\mu\nu}$ in a manner similar to that discussed in Ref.~\cite{BC94},
\beqa
\eta_{\mu\nu}&\leq& \frac{1}{2}\sum_{n\neq k}(p^{0}_{n}+p^{0}_{k})\langle \partial_{\mu}n|k\rangle\langle k|\partial_{\nu}n\rangle, \nonumber \\
&=&\sum_{n\neq k}p^{0}_{n}\langle \partial_{\mu}n|k\rangle\langle k|\partial_{\nu}n\rangle, \nonumber \\
&=&\sum_{n}p^{0}_{n}\langle \partial_{\mu}n|(1-|n\rangle\langle n|)|\partial_{\nu}n\rangle, \nonumber \\
&=&\sum_{n}p^{0}_{n}g^{(n)}_{\mu\nu}.
\eeqa
As a result, it immediately follows that
\beq
\int_{\lambda_{0}}^{\lambda_{\tau}}\sqrt{\eta_{\mu\nu}d\vl^{\mu}d\vl^{\nu}}\leq \int_{\lambda_{0}}^{\lambda_{\tau}}\sqrt{\sum_{n}p^{0}_{n}g^{(n)}_{\mu\nu}d\vl^{\mu}d\vl^{\nu}},
\eeq
which completes the proof of inequality~(\ref{supp:ineq}).

Next, we discuss some properties of the distance $\ell(\rho(0),\rho(\tau))$. Using the concavity of the square root, we note that
the distance $\ell(\rho(0),\rho(\tau))$ can be lower-bounded in terms of the ensemble average of the fidelity of each mode as 
\beqa
\ell(\rho(0),\rho(\tau))
&:=&\int^{\lambda_{\tau}}_{\lambda_{0}}\sqrt{\sum_{n}p_n^0 g_{\mu\nu}^{(n)}d{\lambda}^\mu d{\lambda}^{\nu}},\nonumber \\
&\geq&\sum_{n}p_n^0\int^{\lambda_{\tau}}_{\lambda_{0}} \sqrt{g_{\mu\nu}^{(n)}d{\lambda}^\mu d{\lambda}^{\nu}},\nonumber \\
&\geq &\sum_{n}p_n^0\arccos \sqrt{F(|n(0)\ra,|n(\tau\ra)},
\eeqa
where the fidelity between the $n$-th energy eigenstates at time $t=0$ and $\tau$ takes the form
\beqa
F(|n(0)\ra,|n(\tau\ra)
&=&|\la n(0)|n(\tau\ra|^2.
\eeqa

If we consider the zero temperature limit, i.e., $\rho(\lambda_{0})=|\psi(0)\ra\la \psi(0)|$ and $\rho(\tau)=|\psi(\tau)\ra\la \psi(\tau)|$ with $|\psi\ra:=|0\ra$ denoting the ground state, we have
\beqa
\ell(|\psi(0)\ra,|\psi(\tau\ra)&=&\int^{\lambda_{\tau}}_{\lambda_{0}}\sqrt{ g_{\mu\nu}^{(0)}d{\lambda}^\mu d{\lambda}^{\nu}}, \nonumber \\
&\geq&\arccos \sqrt{F(|\psi(0)\ra,|\psi(\tau\ra)}, \nonumber \\
&= &\mathcal{L}(|\psi(0)\ra,|\psi(\tau\ra), \label{supp:zerot}
\eeqa
and the distance $\ell$ reduces to the Bures length for pure states if the protocol $\lambda_{t}$ follows the geodesic, i.e.,  $\int^{\lambda_{\tau}}_{\lambda_{0}}\sqrt{ g_{\mu\nu}^{(0)}d{\lambda}^\mu d{\lambda}^{\nu}}$ is minimized. 

We finally show the inequality
\beq
 \delta (\Delta W)^{2} \leq (\Delta E_{\rm CD})^{2},\label{supp:we}
\eeq
this is, that the excess of work fluctuations during CD is bounded from above by the energy fluctuations.
From Eqs.~(\ref{keyeq4}) and (\ref{wvarq}), we have
\beqa
\delta (\Delta W)^{2}&=&\sum_{n,m}p^{0}_{n}p_{n\rightarrow m}^t \biggl[(E_m (t)-\varepsilon_n (0))^2-(\varepsilon_n (t)-\varepsilon_n (0))^2\biggr] \nonumber \\
&=&\sum_{n,m}p^{0}_{n}p_{n\rightarrow m}^t \biggl[E_m^{2} (t)-\varepsilon_n^{2} (t)\biggr]\nonumber \\
& &-2\sum_{n,m}p^{0}_{n}p_{n\rightarrow m}^{t} \varepsilon_{n}(0)(E_{m}(t)-\varepsilon_{n}(t))\nonumber \\
&=&\la H^{2}_{\rm CD}\ra-\la H^{2}_{0}\ra,
\eeqa
where the term in the third line vanishes by using Eq.~(\ref{keyeq3}). Noting that 
\beq
\la H^{2}_{0}\ra \geq \la H_{0}\ra^{2}=\la H_{\rm CD}\ra^{2},
\eeq
we obtain~(\ref{supp:we}).


\section{Ion-trap realization}

\begin{figure}[t]
\includegraphics[width=2.7in]{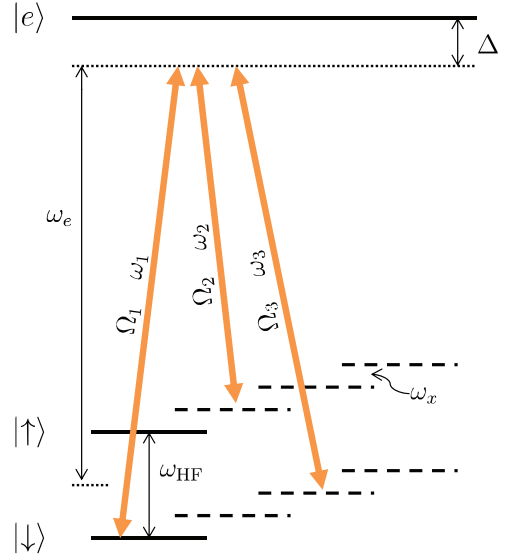}
\caption{Scheme of the energy spectrum of an $^{171}{\rm Y_b}^+$ ion confined in a linear Paul trap. The effective laser-induced potential and the counterdiabatic term are generated by three copropagating laser beams with ${\mathbf k}_1=-{\mathbf k}_2=-{\mathbf k_3}\equiv{\mathbf k}$.\label{fig:EnergyLevels.}}
\label{FigSM1}
\end{figure}

In this section, we show in detail how the use of  two Raman processes  induces an effective Harmonic oscillator with time-dependent frequency as well as the counterdiabatic driving term. We consider the configuration displayed in  Fig. \ref{FigSM1}. The bare Hamiltonian of a trapped $^{171}{\rm Yb}^+$ ion interacting with three Raman laser beams is written as follows,
\begin{eqnarray}
\hat H\left(t\right)&=&\frac{\hbar\omega_{\rm hf}}{2}\hat\sigma_z+\hbar\omega_e\left|e\right\rangle\left\langle e\right|+\frac{\hat p^2}{2m}+\frac{1}{2}m\omega_0^2\hat q^2\\
&&+\sum_{i=1,3}\hbar\Omega_i\left(t\right)\cos\left(k_i\hat q-\omega_it+\varphi_i\right)\left(\left|e\right\rangle\left\langle\downarrow\right|+\left|\downarrow\right\rangle\left\langle e\right|\right)\nonumber\\
&&+\hbar\Omega_2\left(t\right)\cos\left(k_2\hat q-\omega_2t\right)\left(\left|e\right\rangle\left\langle\uparrow\right|+\left|\uparrow\right\rangle\left\langle e\right|\right),\nonumber
\end{eqnarray}
where $\hbar\omega_{\rm hf}$ is the energy splitting between the up state $\left|\uparrow\right\rangle$ and the down state $\left|\downarrow\right\rangle$, with the corresponding Pauli matrices $\hat\sigma_{x,y,z}$, and $\hbar\omega_e$ being the energy of the excited state $\left|e\right\rangle$. The trapping frequency is denoted by $\omega_0$. In addition, the frequencies,  wave vectors,  time-dependent Rabi couplings and the phase of the Raman beams are denoted by $\omega_i$, $k_i$ $\Omega_i\left(t\right)$ and $\varphi_i$, respectively.

We define the ladder operators of the harmonic oscillator as 
\begin{eqnarray}
\hat a=\sqrt{\frac{m\omega_0}{2\hbar}}\left(\hat q+\frac{i\hat p}{m\omega_0}\right),\quad \hat a^\dag=\sqrt{\frac{m\omega_0}{2\hbar}}\left(\hat q-\frac{i\hat p}{m\omega_0}\right).
\end{eqnarray}
In order to induce two-photon Raman process, the frequencies of the three Raman laser beams are chosen to satisfy
\begin{eqnarray}
\omega_1&=&\omega_e+\frac{\omega_{\rm hf}}{2}-\Delta+\left(\omega_0-\nu\right),\\
\omega_2&=&\omega_e-\frac{\omega_{\rm hf}}{2}-\Delta+\delta,\nonumber\\
\omega_3&=&\omega_e+\frac{\omega_{\rm hf}}{2}-\Delta-\left(\omega_0-\nu\right),\nonumber
\end{eqnarray}
where the detuning $\Delta$ to the excited state should be large enough to prevent population transfer to the excited state.
Under the optical rotating-wave approximation, the Hamiltonian in the interaction picture defined by $\hat H_0=\frac{\hbar\omega_{\rm hf}}{2}\hat\sigma_z+\hbar\omega_e\left|e\right\rangle\left\langle e\right|+\hbar\omega_0\left(\hat a^\dag\hat a+\frac{1}{2}\right)$ reads
\begin{eqnarray}
\hat H_I\left(t\right)&=&\sum_{i=1,3}\frac{\hbar\Omega_i\left(t\right)}{2}\left(\exp\left[i\eta_i\left(\hat ae^{-i\omega_0t}+\hat a^\dag e^{i\omega_0t}\right)\right]e^{-i(\omega_it-\varphi_i)}+{\rm h.c.}\right)\nonumber\\
&&\times\left(\left|e\right\rangle\left\langle\downarrow\right|e^{i\left(\omega_e+\omega_{\rm hf}/2\right)t}+\left|\downarrow\right\rangle\left\langle e\right|e^{-i\left(\omega_e+\omega_{\rm hf}/2\right)t}\right)\nonumber\\
&&+\frac{\hbar\Omega_2\left(t\right)}{2}\left(\exp\left[i\eta_2\left(\hat ae^{-i\omega_0t}+\hat a^\dag e^{i\omega_0t}\right)\right]e^{-i\omega_2t}+{\rm h.c.}\right)\nonumber\\
&&\times\left(\left|e\right\rangle\left\langle\uparrow\right|e^{i\left(\omega_e-\omega_{\rm hf}/2\right)t}+\left|\uparrow\right\rangle\left\langle e\right|e^{-i\left(\omega_e-\omega_{\rm hf}/2\right)t}\right),
\end{eqnarray}
where $\eta_i$ denotes the Lamb-Dicke parameter for the corresponding laser, and $\eta_1=-\eta_2=-\eta_3=\eta$ as it follows from the wave vectors relation.

After adiabatically eliminating the excited state $\left|e\right\rangle$ with $\varphi_{1,2}=0$, we obtain the following effective Hamiltonian,
\begin{eqnarray}
\tilde H_{0}^{\rm eff}\left(t\right)&=&\frac{\hbar\eta^2\Omega_1\left(t\right)\Omega_3\left(t\right)}{2\Delta}\left(\hat a\hat a e^{-i(2\nu t+\varphi_{3})}+\hat a^\dag\hat a^\dag e^{i(2\nu t+\varphi_{3})}\right)\left|\downarrow\right\rangle\left\langle\downarrow\right| \nonumber \\
&+&\frac{i\hbar\eta\Omega_1\left(t\right)\Omega_2\left(t\right)}{2\Delta}\left(\hat a\left|\downarrow\right\rangle\left\langle\uparrow\right|e^{-i\left(\nu+\delta\right)t}-\hat a^\dag\left|\uparrow\right\rangle\left\langle\downarrow\right|e^{i\left(\nu+\delta\right)t}\right).\nonumber
\end{eqnarray}
We next consider the constraint  $\delta\gg{\rm max}\left[\frac{\eta\Omega_1\left(t\right)\Omega_2\left(t\right)}{\Delta}\right]$, and adiabatically eliminate the up state $\left|\uparrow\right\rangle$. The effective Hamiltonian after this step becomes
\begin{eqnarray}
\tilde H_{0}^{\rm eff}\left(t\right)&=&\frac{\hbar\eta^2\Omega_1\left(t\right)\Omega_3\left(t\right)}{2\Delta}\left(\hat a\hat a e^{-i(2\nu t+\varphi_{3})}+\hat a^\dag\hat a^\dag e^{i(2\nu t+\varphi_{3})}\right) \nonumber \\
&-&\frac{\hbar\eta^2\Omega_1^2\left(t\right)\Omega_2^2\left(t\right)}{8\left(\nu+\delta\right)\Delta^2}\left(\hat a\hat a^\dag+\hat a^\dag\hat a\right)+{\rm A.C.},
\end{eqnarray}
where A.C. $=\frac{\hbar\eta^2\Omega_1^2\left(t\right)\Omega_2^2\left(t\right)}{8\left(\nu+\delta\right)\Delta^2}$ is the light shift by the Raman laser beams. This term  introduces only a global phase that can be gauged away and we shall ignore it in the rest of the supplementary information.  
We next make use of a further unitary transformation defined by $\hat U\left(t\right)=\exp\left[i\nu t\left(\hat a^\dag\hat a+\frac{1}{2}\right)\right]$ to obtain the following effective Hamiltonian,
\begin{eqnarray}
\hat H_{\rm eff}\left(t\right)&=&i\hbar\frac{\partial\hat U}{\partial t}+\hat U^\dag\tilde H_{0}^{\rm eff}\left(t\right)\hat U\left(t\right) \nonumber \\
&=&\hbar\nu\left(\hat a^\dag\hat a+\frac{1}{2}\right)+\frac{\hbar}{2}\left(\Omega_{{\rm eff},1}\left(t\right)\hat a\hat a+\Omega^{*}_{{\rm eff},1}\left(t\right)\hat a^\dag\hat a^\dag\right)\nonumber\\
&-&\frac{\hbar\Omega_{{\rm eff},2}\left(t\right)}{2}\left(\hat a\hat a^\dag+\hat a^\dag\hat a\right),\label{effectiveH}
\end{eqnarray}
with
\begin{eqnarray*}
\Omega_{{\rm eff},1}\left(t\right)=\frac{\eta^2\Omega_1\left(t\right)\Omega_3\left(t\right)e^{-i\varphi_{3}}}{\Delta},\quad \Omega_{{\rm eff},2}=\frac{\eta^2\Omega_1^2\left(t\right)\Omega_2^2\left(t\right)}{4\left(\nu+\delta\right)\Delta^2}.
\end{eqnarray*}

To introduce an effective time-dependent trapping frequency, we adjust the laser's coupling strengths $\Omega_i\left(t\right)$ and take the phase $\varphi_{3}=0$ to satisfy 
\begin{eqnarray}
\Omega_{{\rm eff},1}\left(t\right)=-\Omega_{{\rm eff},2}\left(t\right)=-\Omega\left(t\right),\label{Raman}
\end{eqnarray}
where $\Omega\left(t\right)$ is a positive real function. The effective Hamiltonian can be rewritten in the first quantized form as 
\begin{eqnarray}
\hat H_{\rm eff}\left(t\right)&=&\hbar \nu \left(\hat a^\dag\hat a+\frac{1}{2}\right)-\frac{\hbar \Omega(t)}{2}\left(\hat{a}+\hat{a}^{\dagger}\right)^{2}, \nonumber \\
&=&\frac{\nu}{m\omega_{0}}\hat{p}^{2}+\frac{m\omega_{0}\nu}{2}\hat{q}^{2}-m\omega_{0}\Omega(t)\hat{q}^{2},\nonumber \\
&=&\frac{\hat p^2}{2m_{\rm eff}}+\frac{1}{2}m_{\rm eff}\omega\left(t\right)^2\hat q^2,
\end{eqnarray}
where $m_{\rm eff}=m\omega_0/\nu$ is the effective mass and $\omega\left(t\right)=\sqrt{\nu\left(\nu-\Omega\left(t\right)\right)}$. Note that the Hamiltonian above is equivalent to a harmonic oscillator with a time-dependent trapping frequency $\omega(t)$.

The counterdiabatic field $\hat H_1\left(t\right)$ for a time-dependent harmonic oscillator can be written as
\begin{eqnarray}
\hat H_1\left(t\right)&=&-\frac{\dot\omega\left(t\right)}{\omega\left(t\right)}\left(\hat q\hat p+\hat p\hat q\right)\nonumber \\
&=&\frac{i\hbar\dot\omega\left(t\right)}{4\omega\left(t\right)}\left(\hat a\hat a-\hat a^\dag\hat a^\dag\right).\label{CF}
\end{eqnarray}
Therefore, if we further tune the relative phase $\varphi_{3}$ of the Raman laser as
\begin{eqnarray}
\Omega_{{\rm eff},1}\left(t\right)=-\Omega\left(t\right)+\frac{i}{2}\frac{\dot\omega\left(t\right)}{\omega\left(t\right)},\quad\Omega_{{\rm eff},2}=\Omega\left(t\right),
\end{eqnarray}
the effective Hamiltonian~(\ref{effectiveH}) becomes
\begin{eqnarray}
\hat H_{\rm eff}\left(t\right)=\frac{\hat p^2}{2m_{\rm eff}}+\frac{1}{2}m_{\rm eff}\omega\left(t\right)^2\hat q^2-\frac{\dot\omega\left(t\right)}{\omega\left(t\right)}\left(\hat q\hat p+\hat p\hat q\right),
\end{eqnarray}
which is equal to the counterdiabatic Hamiltonian $\hat{H}_{\mathrm{CD}}[\omega(t)]$ of the time-dependent Harmonic oscillator including the counterdiabatic field~(\ref{CF}). The full-driving Hamiltonian $\hat{H}_{\rm eff}(t)=\hat{H}_{\mathrm{CD}}[\omega(t)]$ takes the form of a generalized time-dependent harmonic oscillator. The instantaneous eigenenergies of $\hat{H}_{\rm CD}$ are given by
\beqa
E_n=\hbar\om\sqrt{1-\frac{\dot{\om}^2}{4\om^4}}\left(n+\frac{1}{2}\right),
\eeqa
and the  corresponding eigenfunctions are given by 
\beqa
& & \psi_n(q)=\frac{1}{\sqrt{2^n n!}}\left(\frac{m\om}{\hbar\pi}\sqrt{1-\frac{\dot{\om}^2}{4\om^4}}\right)^{\frac{1}{4}}
\!\!\!H_n\bigg[\sqrt{\frac{m\om}{\hbar}\sqrt{1-\frac{\dot{\om}^2}{4\om^4}}}q\bigg]\nonumber\\
& &\times 
\exp\biggl[
-\frac{m\om}{2\hbar}\sqrt{1-\frac{\dot{\om}^2}{4\om^4}} q^{2}\biggr]
\exp\biggl[i\frac{m\dot{\om}}{4\hbar\om}q^{2}\biggr],
\eeqa
where $\om(t)$ is to be considered as the control parameter $\lambda(t)$ of the Hamiltonian. 

\end{document}